# Rapid Trapping and Label-free Characterization of Single Nanoscale Extracellular Vesicles and Nanoparticles in Solution

Ikjun Hong[1,2], Chuchuan Hong[1,2], Theodore Anyika[1,2], Guodong Zhu[1,2], and Maxwell Ugwu[1,3], Jeff Franklin[4,5,7], Robert Coffey[4,5,7,8], Justus C. Ndukaife[1,2,6,7]*
*Correspondence: justus.ndukaife@vanderbilt.edu

[1]Vanderbilt Institute of Nanoscale Science and Engineering, Vanderbilt University, Nashville, Tennessee, 37235, United States

[2]Department of Electrical and Computer Engineering, Vanderbilt University, Nashville, Tennessee, 37235, United States

[3]Interdisciplinary Materials Science and Engineering, Vanderbilt University, Nashville, Tennessee, 37235, United States

[4]Department of Medicine, Vanderbilt University Medical Center, Nashville, Tennessee 37232, United States

[5]Department of Cell and Developmental Biology, Vanderbilt University, Nashville, Tennessee 37232, United States

[6]Department of Mechanical Engineering, Vanderbilt University, Nashville, Tennessee 37235, United States

[7]Center for Extracellular Vesicles Research, Vanderbilt University, Nashville, Tennessee 37235, United States

[8]Epithelial Biology Center, Vanderbilt University Medical Center, Nashville, TN

## Abstract:

Achieving high-throughput, comprehensive analysis of single nanoparticles to determine their size, shape, and composition is essential for understanding particle heterogeneity with applications ranging from drug delivery to environmental monitoring. Existing techniques are hindered by low throughput, lengthy trapping times, irreversible particle adsorption, or limited characterization capabilities. Here, we introduce Interferometric Electrohydrodynamic Tweezers (IET), an integrated platform that rapidly traps single nanoparticles in parallel within three seconds. IET enables label-free characterization of particle size and shape via interferometric imaging and identifies molecular composition through Raman spectroscopy, all without the need for fluorescent labeling. We demonstrate the platform's capabilities by trapping and imaging colloidal polymer beads, nanoscale extracellular vesicles (EVs), and newly discovered extracellular nanoparticles known as supermeres. By monitoring their interferometric contrast images while trapped, we accurately determine the sizes of EVs and supermeres. Our IET represents a powerful optofluidics platform for comprehensive characterization of nanoscale objects, opening new avenues in nanomedicine, environmental monitoring, and beyond.

**Introduction:**

Understanding the heterogeneity of nanoscale biological objects, such as extracellular vesicles (EVs) and newly discovered non-vesicular extracellular nanoparticles[1], is essential for advances in nanomedicine[2], diagnostics[3], and environmental monitoring[3,4]. However, current methods for manipulating, imaging, and analyzing these nanoscale objects are constrained by low throughput, complex workflows, or limited characterization capabilities. Techniques such as anti-Brownian electrokinetic trapping [5,6], plasmonic nanotweezers [7–14], resonant dielectric nanoantenna traps [15,16], quasi-BIC metasurface tweezers[17–23], optothermoelectric tweezer[24], and photonic crystal cavity traps [25–27] have shown promise in trapping nanoparticles, but fall short of delivering the high-throughput, label-free, and comprehensive analysis required to unlock the full potential of nanoparticle heterogeneity studies.

Label-free imaging technologies, such as interferometric scattering (iSCAT) and coherent bright-field imaging (COBRI)[28], have emerged as powerful tools for visualizing nanoscale objects without perturbing their native states[29–31]. These methods rely on the interaction between a reference beam and the Rayleigh-scattered signal from subwavelength-scale targets, allowing for fast imaging and bypassing the limitations of fluorescence-based techniques, such as photobleaching and labeling artifacts. In iSCAT microscopy, the backward-scattered light is interfered with a portion of the incident light reflected at the water-substrate interface. In contrast, COBRI interferes the forward-scattered light with the transmitted incident light. Both iSCAT and COBRI have enabled the study of particle dynamics with unprecedented temporal resolutions. However, existing implementations often require particles to diffuse to and bind irreversibly to surfaces, introducing variability in particle analysis (since particles of different sizes diffuse over different time scales) and limiting reusability. For example, in the single-particle interferometric reflectance imaging sensor (SP-IRIS)[32], particles are irreversibly captured on antibody-functionalized surfaces, necessitating new samples for each experiment.

Despite the advances in single particle trapping and characterization, we note that a scalable platform that offers simultaneous high-throughput trapping, label-free imaging, and molecular composition analysis of individual nanoparticles remains elusive. Such a tool would revolutionize single-particle analysis by enabling rapid, detailed characterization of both size and chemical composition at the nanoscale, with profound implications across a range of fields, from nanomedicine to environmental science.

To address this critical gap, we introduce an original interferometric electrohydrodynamic tweezers (IET). IET uses electrohydrodynamic flows to rapidly trap thousands of nanoscale objects, such as EVs and nanoparticles, in parallel—within seconds. Our platform integrates label-free interferometric imaging and molecular composition analysis using Raman spectroscopy, enabling precise, real-time characterization at the single particle level without the need for fluorescent labels or surface immobilization. Importantly, IET allows for the comprehensive analysis of nanoscale particles (including size, shape and chemical compostion) in their native state, avoiding artifacts introduced by traditional staining or fixation techniques.

We demonstrate the versatility and power of IET by achieving rapid trapping and label-free imaging of colloidal polymer beads, EVs, and newly discovered extracellular nanoparticles, including supermeres, with unprecedented speed. By monitoring the interferometric contrast images, we determine the size of these particles while they remain freely suspended, offering a unique advantage over conventional surface-bound techniques.

Our IET platform represents a transformative tool for nanoscale particle trapping and characterization, offering new possibilities for high-throughput analysis of biological and synthetic nanoparticles, with broad applications in fields such as nanomedicine, drug delivery, and environmental monitoring.

## Working principle of interferometric electrohydrodynamic tweezers and experimental set-up

Our IET platform illustrated in Figure 1a is comprised of a gold film patterned with an array of micron scale holes, which can be readily fabricated via photolithography to enable scalable wafer-scale large area samples as desired. A bottom electrode made of materials such as ITO is placed on top of the patterned gold film with a dielectric spacer layer in between and the chip is mounted on a custom microscope set-up as depicted in Figure 1b. The dielectric spacer layer forms the microfluidic channel where the micron scale hole array and the fluidic medium is contained.

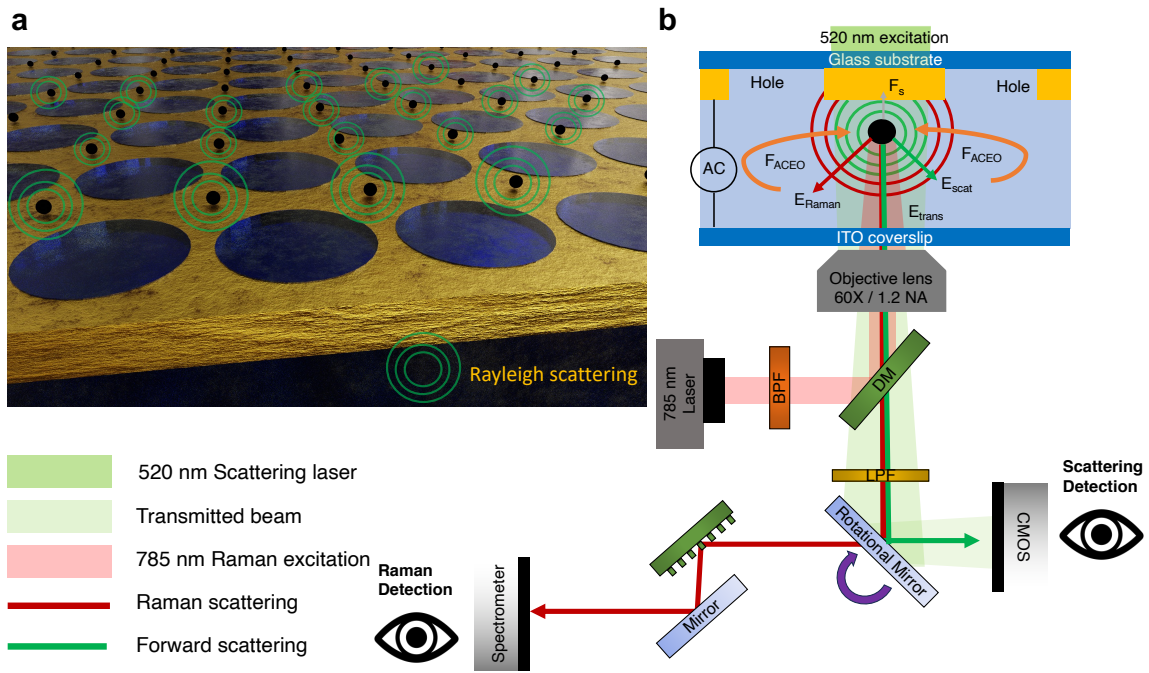

**Figure 1. Working principle of the Interferometric Electrohydrodynamic Tweezers (IET) system.**
**a,** Schematic view of the IET system capable of rapid parallel trapping of multiple particles and label-free interferometric imaging. **b,** Illustration of the working principle and experimental setup of the IET system. Nanoparticles are rapidly transported by in-plane AC electro-osmotic (ACEO) flow and trapped at the nearest stagnation zones. Forces acting on the particles include $F_{ACEO}$, the drag force from ACEO flow, and Fs, the particle surface interaction force. Optical signals involved are Escat, the scattered electric field from the particle; ERaman, the Raman signal from the particle; and Etrans, the transmitted light after passing through the thin gold film. Optical components include: DM, the dichroic mirror; BPF, the band-pass filter; and LPF, the long pass filter.

When an alternating a.c. voltage with frequency ranging from 1 kHz to 10 kHz is applied across the patterned gold film, both normal and tangential a.c. electric fields are generated. The tangential a.c. electric field acts on the electric double layer (EDL) charges induced at the interface between the gold film and the adjoining fluid medium to produce surface ACEO flows, the flow velocity of which is defined by the Helmholtz-Smoluchowski slip velocity with magnitude that is given by[33–35]:

$$u_s = -\frac{\varepsilon_w \zeta}{\eta} E_\parallel,$$

where $u_s$ is the velocity of the ACEO flow, $\varepsilon_w$ is the permittivity of the fluid medium, $\zeta$ is the zeta potential, $\eta$ is the fluid viscosity, and $E_\parallel$ is the tangential component of the electric field established near the gold-fluid interface. More details on the numerical simulations are provided in Supplementary Information section 1. Interestingly, the induced ACEO flow field vectors converge at the center of the unpatterned regions of the gold film to form a stagnation zone where the in-plane fluid velocity goes to zero as depicted in Figure 2a. These stagnation zones provide the regions where single nanoscale objects can be readily localized. The localization of the particle in the out-of-plane direction is mediated by the particle-surface interaction force that arises from the interaction between the double layer charge on the particle and its image charge in the conduction plane[36].

Once the particles have been trapped at the respective trapping sites, they can be readily imaged without attaching them to the gold surface or any fluorescent labels by using interferometric imaging. To achieve this, an imaging laser beam of 520 nm in wavelength illuminates the sample from the top so that some of the light is transmitted through the gold film. A portion of the light transmitted through the gold film is scattered by the trapped particles before reaching the detector. The detected light reaching the detector, given by the transmitted light and the scattered light is described as follows[37]:

$$I_{det} = |E_{tran} + E_{scat}|^2 = |E_{inc}|^2(t^2 + |s|^2 + 2t|s|\cos\theta),$$

where $E_{tran}$ is the transmitted electric field, $E_{scat}$ is the scattered electric field, $t$ is the transmission coefficient, and $|s|^2$ is scattering cross-section from the particle in the medium, and $\theta$ is the phase difference between the scattered light and the light transmitted through the thin gold film. The intensity normalization between the detected and incident transmitted light, which is $\frac{I_{det}}{I_{tran}}$, yields the contrast after the imaging process. The contrast term can be described as

$$C = \left(\frac{|s|}{t}\right)^2 + 2\frac{|s|}{t} = \left(\frac{1}{t}\right)^2 \sigma_{scat} + 2\left(\frac{1}{t}\right)\sqrt{\sigma_{scat}},$$

where the $\sigma_{scat}$ is the scattering cross-section from the particle.

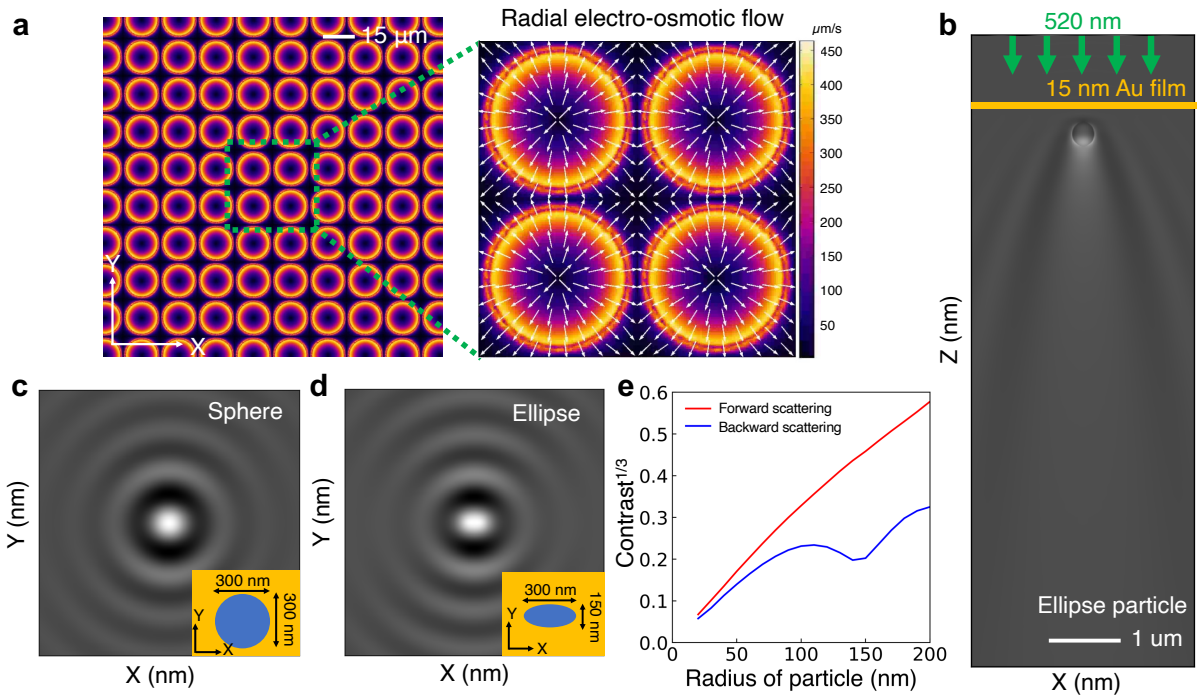

**Figure 2. Working principle of interferometric electrohydrodynamic tweezers (IET) and scattering simulations on 300 nm spherical and elliptical nanoparticles. a**, Electrohydrodynamic simulation showing that the IET-induced radial ACEO flow field converge radially to establish stagnation zones at the center of four adjacent holes where the nanoparticles can be trapped. **b**, Finite-Difference Time-Domain (FDTD) simulation depicting the intensity of scattered light along the out-of-plane direction, normalized by the background field transmitted through a 15 nm thin gold film. **c**, **d**, Intensity of scattered light, normalized by the background field along the in-plane direction, illustrating different scattering patterns for nanoparticles of different shapes: **c,** a 300 nm diameter spherical particle, and **d,** an elliptical particle with a length of 150 nm along the Y-axis. Insets show the cross-section of particle shapes. **e,** Simulated contrast of forward and backward scattering as a function of particle size, indicating that forward Mie scattering becomes dominant as particle size increases. This trend is advantageous for the IET platform, which collects forward-scattered photons, allowing to correlate particle size with contrast even for larger particles.

Here, the forward scattered light results in an enhanced image contrast after removing the background transmitted light as depicted in Figure 2b (see Methods section). Figure 2c and d shows the interferometric contrast image for particles of varying shapes ranging from spherical to ellipsoidal shape. The results show that the shape of the particle can also be detected based on their interferometric image in the IET chip.

We note that interferometric scattering (iSCAT) experiments typically operate in reflection mode, where particles or molecules are placed on a nearly transparent substrate, such as glass with weak reflectivity[3,30,38–40]. In this setup, the detector collects the backscattered light from the adsorbed particles. It has been observed that the contrast achieved by collecting the backscattered photons does not follow a linear relationship with particle size, as shown by the blue curve of Figure 2e[37,41]. In contrast, our IET platform operates in transmission mode, where the forward-scattered light is collected by the detector, similar to coherent bright-field imaging[42]. We find that the contrast associated with the forward-scattered light exhibits a nearly linear relationship with particle size, as shown by the red curve of Figure 2e (see Methods section). This occurs because as the particle size increases and approaches half the wavelength, forward scattering dominates over backward scattering. By operating in the transmission mode, IET platform ensures that the predominant forward scattered photons are collected. Therefore, the size of trapped particles can be detected and characterized using the IET platform by analyzing the interferometric contrast signal (details provided in the subsequent experimental section). Subsequently, the Raman signal from a trapped nanoparticle can be readily collected to identify the chemical composition.

**Particle trapping and label-free characterization**

We experimentally demonstrated the rapid parallel trapping and label-free imaging of extracellular vesicles (EVs), supermeres, and dielectric beads using the IET platform. The platform, fabricated using photolithography (detailed in the Methods/Supplementary Information section 2), spans a large area of 8 mm by 4 mm and contains an array of 60,000 trapping sites. These sites are formed by 15 µm diameter holes patterned with a periodicity of 18 µm, as depicted in the SEM image in the Supplementary Information section 3. The IET platform employs a 15 nm-thick gold (Au) film, chosen for its sufficient electrical conductivity to establish tangential electric fields and in-plane ACEO flow. This film thickness also ensures that the film is transparent to provide sufficient transmitted photons that can be scattered by the trapped nanoscale particles. The chip was fabricated using a photolithography approach to create large-scale patterns.

To evaluate the trapping and label-free imaging in the IET system, initially, 300 nm dielectric beads were introduced into a microfluidic channel embedded with the IET patterns. An AC voltage of 10 V at a frequency of 3 kHz was applied, generating thousands of ACEO flows near the respective trapping sites in parallel. These flows rapidly transported and trapped particles at each of the trapping sites within three seconds (Supplementary Video 1 and 2), as illustrated in Figure 3.

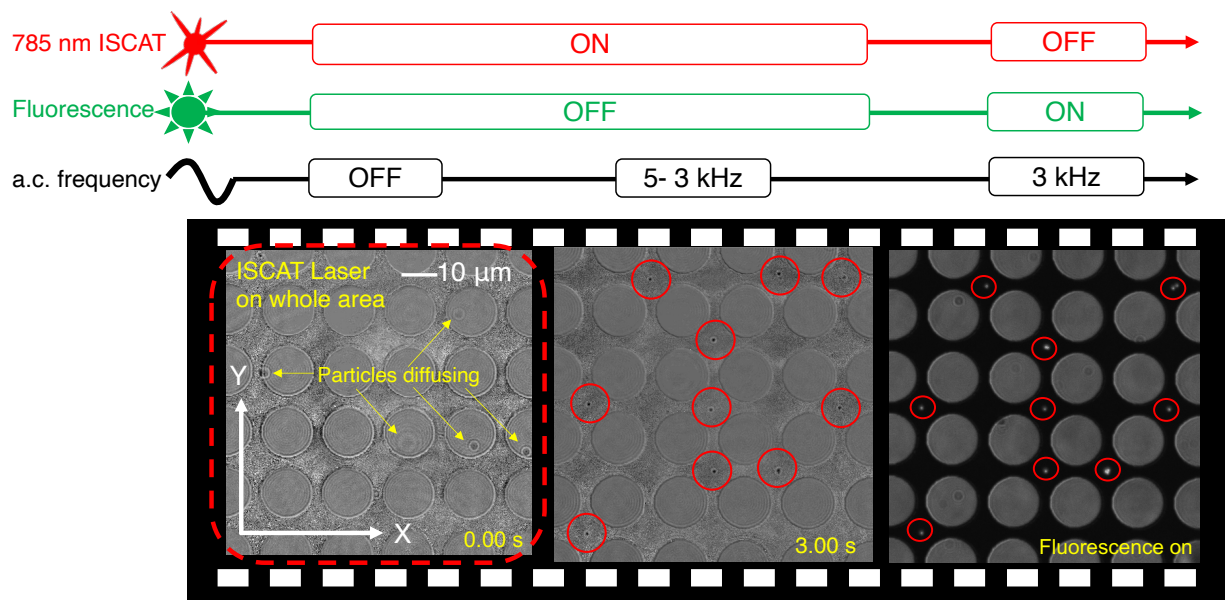

**Figure 3. Rapid parallel trapping and label-free imaging of dielectric nanoparticles using Interferometric Electrohydrodynamic Tweezers (IET).** Frame-by-frame images showing the trapping of 300 nm polystyrene (PS) beads in parallel within seconds combined with label-free imaging. The interferometric (label-free) images in IET compares very well with the fluorescent images.

The overall imaging process is outlined in the Supplementary Information 4. First, the acquired video is averaged, by calculating the mean pixel intensity, and the average frame is used to normalize each frame to remove the background. This normalization enhances the contrast between the transmitted light and the light scattered by the trapped particles, allowing the scattering signal from individual nanoscale particles to be detected with greater precision. For a 60X, 1.2 NA objective lens, the field of view (FOV) is 80 µm x 80 µm, which can contain about 25 IET trapping sites within the FOV.

At this point, our system provides the flexibility to perform any or all of the following tasks: (1) record the interferometric contrast of the trapped particles at each site; (2) temporarily release the particles and track their Brownian motion to ascertain the size from their Brownian diffusion and correlate with their interferometric contrast; (3) illuminate a given particle at a given IET trapping site to further stably trap the particle with optical gradient force in addition to the IET trapping potential and collect their Raman signal to identify their molecular composition, without disturbing the particles in the other trapping sites. These options represent an important advancement of IET over prior optical nanotweezers by providing the capability to characterize the size, and chemical composition of trapped particles in a label-free manner in one integrated platform.

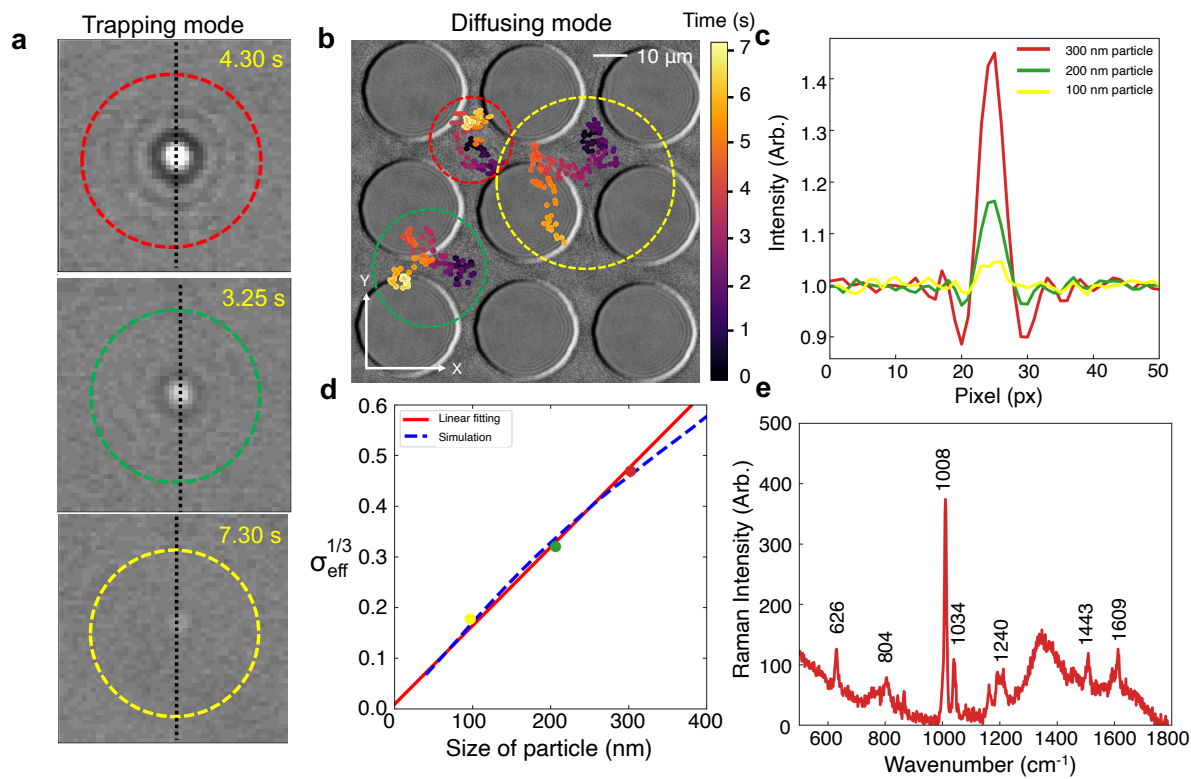

**Figure 4. Comprehensive size and chemical composition analysis of polystyrene (PS) bead particles. a,** The IET platform traps particles of various sizes, and contrast information is extracted while in trapping mode. Red, green, and yellow dotted circles indicate trapped particles with diameters of 300 nm, 200 nm, and 100 nm, respectively. **b,** Releasing the particles by turning off the AC frequency enables analysis of their Brownian motion. Red, green, and yellow dotted circles show the diffusion trajectories of particles after release, and the Brownian diffusion coefficient is calculated through MSD estimation. **c,** Brightness intensity across the black dotted line in the frames of panel a quantifies the contrast for each particle size. **d,** The effective contrast raised to the power of one-third shows a quasi-linear relationship with particle size, marked by a red line. The blue dotted line represents the contrast plot over the particle size from the forward scattering simulation. **e,** Raman measurements of trapped polystyrene particles during IET operation reveal the prominent 1008 cm⁻¹ peak characteristic of polystyrene.

It is evident from Figure 4a, that the interferometric contrast decreases as the particle size decreases. For a scenario where the size of the particles is not known a priori, we note that the size of trapped particles can be determined independently by monitoring their Brownian motion after been released from the IET traps. To elucidate the process of determining the size of the trapped particles in IET, we introduced a mixture of 100 nm, 200 nm, and 300 nm beads. After trapping the particles, the traps are released temporarily by turning off the ac field and we tracked the Brownian motion of the particles (Supplementary Video 3). During the diffusion mode, particles of different sizes undergo Brownian dynamics in the medium, as shown in Figure 4b. The diffusion coefficient is calculated using the equation $D = MSD(\tau)/2d\tau$, where $D$ is the diffusion coefficient, $\tau$ is the lag time, and $d$ is the dimensional factor, which in this case is 2 [43]. By analyzing the diffusion dynamics, we obtained the size information of the trapped particles (see Methods sections). Through the mean square displacement (MSD) calculation detailed in Supplementary Information 5, the slopes of the linear fit of the

MSD points indicate Brownian diffusion coefficients of 5.4 µm²/s, 7.91 µm²/s, and 16.9 µm²/s, corresponding to estimated PS bead diameters of 302.2 nm, 207.2 nm, and 96.6 nm, respectively. These estimated diameters match the predefined sizes of the PS beads, which are 300 nm, 200 nm, and 100 nm in diameter. Additionally, during the trapping phase shown in Figure 4a, contrast information from the nanoparticles is to correlate with the size of the trapped particles. Figure 4c shows the brightness intensity across the center of the Airy disk, marked by a black dotted line in the insets of Figure 4a. Here, the standard deviation of the intensity for the different particles quantifies the contrast for the different particles (see Methods sections)

We proceeded to correlate the size of the particles estimated from Brownian motion with the cube root of their interferometric contrast as shown in Figure 4d. The result in Figure 4d shows that the cube root of the standard deviation of the interferometric contrast scales linearly with particle size, marked by a red line. Also, the contrast calculated by forward-scattering FDTD simulation also closely corresponds to particle size, marked by a blue dotted line (see Methods sections). It thus follows that once a calibration relation has been established, the size of the trapped particles can be readily determined by imaging their interferometric contrast across the IET chip.

Another important capability of the IET platform is the ability to detect the chemical composition of trapped particles. Figure 4e show the Raman spectrum from trapped polystyrene beads obtain by illuminating one of the IET trapping sites with a 785 nm Raman laser beam (Supplementary Information 6, 7 and Supplementary Video 4). The dominant peak at 626 cm$^{-1}$ is characteristic of the C-C stretching mode, the peak at 804 cm$^{-1}$ corresponds to the C-H out-of-plane deformation, the peak at 1008 cm$^{-1}$ is characteristic of the ring breathing mode, the peak at 1034 cm$^{-1}$ represents the in-plane C-H bending of the phenyl rings, the peak at 1443 cm$^{-1}$ corresponds to the C-H bending vibrations, and the peak at 1609 cm$^{-1}$ is characteristic of the C-C stretching vibration in the phenyl ring[44,45].

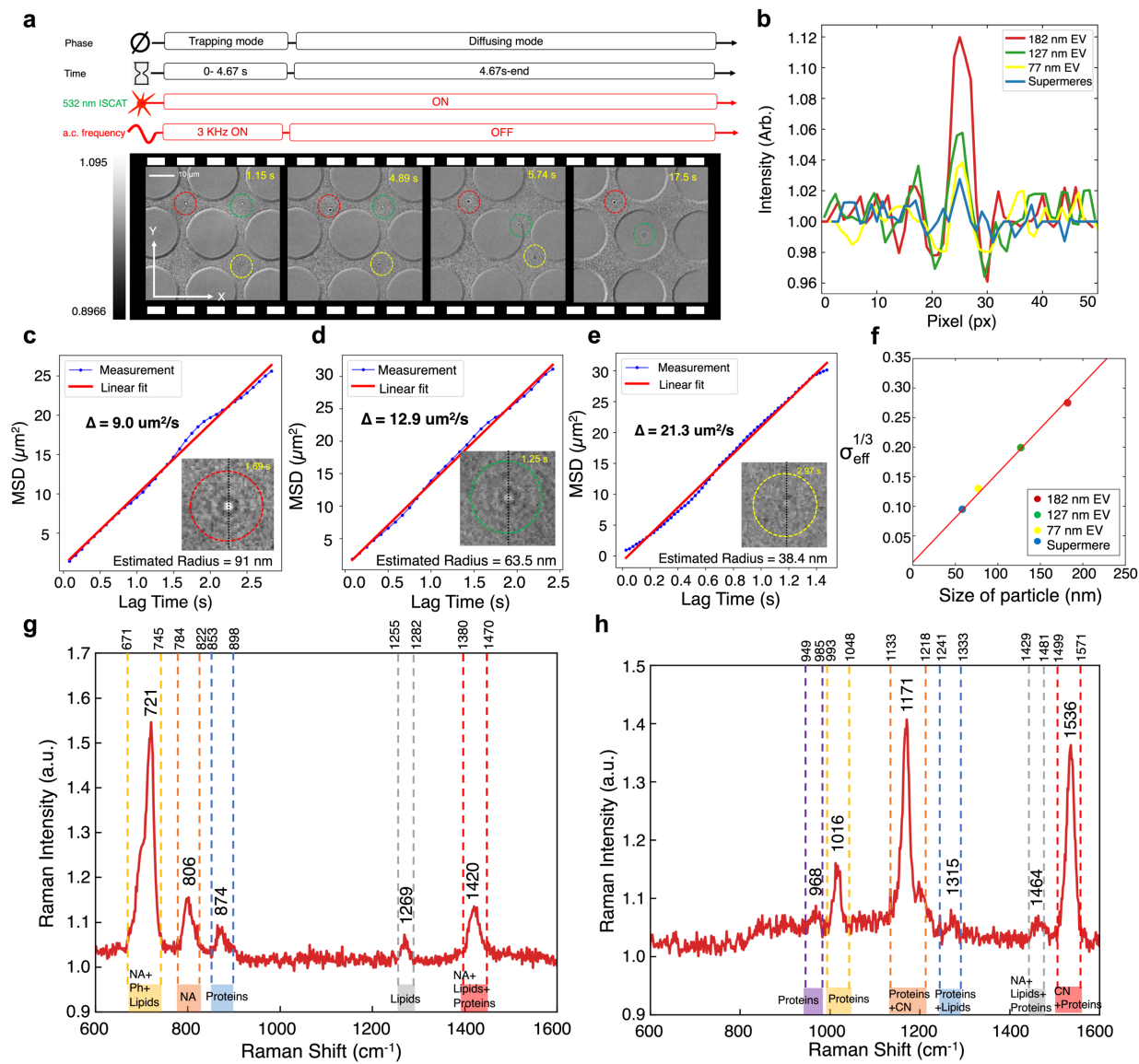

**Figure 5. Label-free imaging and size estimation of extracellular vesicles (EVs) and supermeres. a**, Sequential images showing the trapping and release of EVs in trapping and diffusion modes. **b,** Brightness intensity plot along the black dotted line (inset in panels c, d, e, and Figure S10), showing the relative intensity of trapped EVs and supermeres. **c-e**, Mean square displacement (MSD) calculations based on Brownian motion dynamics, with linear regression indicated by the red line. **f**, The effective contrast raised to the power of one-third exhibits a quasi-linear relationship with the particle size of EVs and supermeres. **g** and **h**, Raman response from the trapped EV1 and EV2 shows distinct Raman peaks.

We also conducted series of experiments to showcase the capability of the IET platform for trapping and label-free imaging of extracellular vesicles and the newly discovered extracellular nanoparticles known as supermeres without the need for any fluorescent labeling. Figure 5a show the frame-by-frame sequence of EV trapping and label-free imaging of the trapped EVs (Supplementary Video 5). Since EVs have heterogenous size distribution, we expect that EVs of varying sizes would be trapped at the respective trapping sites. It is evident that the interferometric contrast images are different for the different EVs sizes trapped at different trapping sites. To independently estimate the EV sizes and generate a calibration plot of their contrast images with respect to size, we released the EVs by temporarily turning OFF the applied AC field and tracking their Brownian dynamics as depicted in Figure S8, similar to the procedure for polystyrene beads.

Figure S8 illustrates the Brownian trajectories of three EVs, with the red, green, and yellow dotted lines tracing their movement in two-dimensional space. Particle localization is tracked over approximately 12 seconds, and the MSD for each EV is shown in Figure 5c, d, and e.

Particles of different sizes display distinct diffusion coefficients in the medium. The slopes ($\Delta$) of the regression lines of MSD calculation for the three cases are 9.0 µm²/s, 12.9 µm²/s, and 21.3 µm²/s, which correspond to diffusion coefficients of 2.25 µm²/s, 3.225 µm²/s, and 5.32 µm²/s, respectively. Based on these diffusion coefficients, the estimated radii of the trapped EVs are 91 nm, 63.5 nm, and 38.4 nm, as shown in Figure 5c, d, and e, which is within the expected size range of small EVs[46,47].

A contrast analysis for the three EVs was also performed. Figure 5b shows the brightness intensity across the center of the Airy disk, marked by a black dotted line in the insets of Figures 5c, d, e, and S10. The contrast is quantified from the standard deviation of the intensity plot across the pixels, as shown in Figure 5f (see in Methods section).

The combination of MSD calculations and contrast analysis during both trapping and diffusion allows for accurate size estimation within our system. We then correlated the particle sizes estimated from Brownian motion with the cube root of their interferometric contrast, as shown in Figure 5f. The results in Figure 5f indicate that the cube root of the standard deviation of the interferometric contrast scales quasi-linearly with EV size. To showcase the capability of the IET platform to profile the global chemical composition of trapped EVs, we also acquired the Raman signals from single EVs that are trapped at their respective trapping sites. Two representative Raman spectra are presented in Figure 5g,h. The Raman spectrum have some differences depicting the heterogeneity of EVs. The trapped EV1, depicted in Fig. 5g, exhibits a prominent Raman response in the range of 671 cm$^{-1}$ to 756 cm$^{-1}$, corresponding to nucleic acids (NA) and phospholipids (Ph), as well as in the range of 1380 cm$^{-1}$ to 1470 cm$^{-1}$, corresponding to NA, lipids, and proteins. Additionally, the other trapped EV2 shows a distinct Raman response in the ranges of 1133 cm$^{-1}$ to 1218 cm$^{-1}$ and 1499 cm$^{-1}$ to 1571 cm$^{-1}$, corresponding to proteins and carotenoids (CN)[48]. The detailed EV-related Raman spectrum and size estimation of the EVs are provided in Supporting Information 11 and 12. The different constituent molecular cargo present in those trapped EVs are indicated in Figure g,h showcasing that the Raman spectroscopy of trapped EVs can provide information on the global biomolecular compositon of the trapped EVs in the IET platform.

It is generally known that EVs have a spherical shape owing to their lipid bilayer configuration[46]. However, when two EVs stick together or if an EV becomes bound to other particles in solution, the overall shape changes. We observe that we can detect if the shape of the trapped EVs deviates from a spherical shape. Specifically, we observed that asymmetric shaped particles exhibit a rotational motion while trapped in the IET which is readily visualized in their interferometric contrast image, as previously computationally studied in Figure 2c and d, and the experimental detail is described in the Supplementary Information 9 and Supplementary Video 6.

Finally, the label-free detection of supermeres using a trapping experiment is also examined (Supplementrary information 10 and Supplementary Video 7). Details of the process for isolating and purifying the supermere samples are provided in the Supplementary Information 13. Figure S10 show the interferometric contrast image of a trapped supermere. From the calibration plot of the cube root of contrast with particle size, we estimated the size of the trapped supermere to be 58 nm from the contrast analysis, marked by a blue dot in Figure 5f.

**Conclusion**

Our interferometric electrohydrodynamic tweezers (IET) platform represents a significant advancement in nanoscale manipulation and characterization, offering unparalleled capabilities for comprehensive analysis of nanoscale objects—including emerging extracellular nanoparticles like supermeres and exomeres. By bridging the gap between high-throughput rapid trapping and label-free imaging, this technology holds immense potential for advancing our understanding of nanometer-scale objects, with broad implications for nanomedicine, environmental monitoring, and beyond.

The IET platform's ability to rapidly trap nanoscale particles—including EVs and newly discovered extracellular nanoparticles—in parallel within seconds, image them in a label-free manner, makes it a promising tool for elucidating the heterogeneity of nanoparticles and EVs. This integrated nanomanipulation and label-free characterization technique represents a significant addition to the toolkit of spectroscopic nanotweezers.

Our approach enables immediate implementation of several exciting applications, such as (1) assessing the purity of EVs and lipoproteins; (2) unraveling EV heterogeneity; (3) characterizing engineered EVs loaded with drug molecules; and (4) nanoplastics characterization. By facilitating comprehensive single-particle analysis without the need for labeling or surface adhesion, the IET platform paves the way for new discoveries and applications in nanoscale science and technology.

**Methods**

**In-plane and Out-of-plane Scattering simulation.** The scattering patterns from the particle, interfered by the reflection from a thin gold film, are studied using commercially available FDTD (Lumerical) analysis. A spherical and elliptical particle are positioned 100 nm above the 15 nm gold film, and an in-plane electric field monitor is placed 25 nm above the top surface of the particle. The boundary conditions along the X, Y, and Z axes are set to perfect matched layer (PML), and plane wave excitation with a x-polarized light is used to analyze the scattering patterns in both the in-plane and out-of-plane directions. First, the background field in the absence of the particle is studied, and the electric field distribution in the presence of the particle is analyzed. Second, each field distribution is squared to obtain the spatial intensity distribution. Third, each intensity distribution is divided accordingly, based on the experimental normalization condition ($\frac{I_{det}}{I_{tran}}$).

**Forward and Backward scattering cross-section simulation.** The contrast between the forward and backward scattering from the particle was calculated using FDTD (Lumerical) analysis. A total-field scattered-field (TFSF) source was used for 520 nm excitation, and a scattering monitor enclosing the TFSF source was placed outside of it to collect only the scattered light from the particle. The particle was positioned 500 nm above the top of a 15 nm thin gold film, while the top scattering monitor was located 500 nm above the particle with a collection area of 2750 nm x 2750 nm to mimic a 70-degree half-angle and a total 140-degree collection angle. Perfectly matched layers were set for the 3D boundary condition, enclosing the entire FDTD calculation domain.

The scattering efficiency, scaled by the scattering cross-section, was then calculated by varying the particle size from 50 nm to 400 nm. Next, the contrast, defined by $C = \left(\frac{1}{t}\right)^2 \sigma_{scat} + 2\left(\frac{1}{t}\right)\sqrt{\sigma_{scat}}$, can be expressed as follows: $C = \beta^2 \sigma_{scat} + 2\beta\sqrt{\sigma_{scat}}$, where $\beta$ is an experimental constant or setup parameter, which depends on the transmission of gold film and ITO-coated glass within the microfluidic channel.

**Fabrication.** The process begins with a $SiO_2$ substrate. First, a 10-second descum process is performed, followed by spin-coating the NR9 photoresist onto the substrate. The coated substrate is then prebaked at 150°C for 60 seconds. Next, the substrate is exposed to 408 nm UV light for 7 seconds through a Cr mask. This step is followed by a post-bake at 100°C for 180 seconds. The exposed areas of the NR9 photoresist are developed using MF-319 for 6 seconds, followed by a 10-second DESCUM process, leaving NR9 pillars on the substrate. Subsequently, a 5 nm Cr adhesion layer is deposited, followed by Au deposition via electron beam evaporation. The final step involves sonication in acetone for 8 minutes to remove the remaining photoresist, resulting in the finished chip.

**MSD calculation.** Tracking the diffusion of particles in two dimensions with the mean-squared-displacement (MSD) of their trajectories provides the diffusion coefficients for each single particle, following the equation: $MSD(\tau) = \langle \Delta r(\tau)^2 \rangle = \langle [r(t+\tau) - r(t)]^2 \rangle$, where $\tau$ is the lag time, $t$ is the designated time, and $r(t+\tau)$ is the position of particle at $t + \tau$ time frames. By tracking the diffusion of particles in two dimensions and calculating the MSD of their trajectories, we obtain the diffusion coefficients for each particle.

**Quantification of contrast.** To quantify the contrast for the scattering from the particles, the standard deviation ($\sigma_{ariy\ disk}$) across the center of the particle is considered, and the standard deviation in the absence of a particle ($\sigma_{background}$) is next experimentally calculated. The difference between them, raised to the power of one-third $\sigma_{ariy\ disk}$-$\sigma_{background})^{1/3}$ is finally considered to be the effective standard deviation from the particle ($\sigma_{eff}^{1/3}$).

**Tracking method.** The particle tracking analysis was performed with the assistance of an open-source Python package named Trackpy (soft-matter/trackpy: Trackpy v0.5.0). The recorded video was converted into image sequences, and then the built-in tracking algorithm was used to identify the location of the fluorescence of trapped particles in each image.